\documentclass[12pt,aps,prc,showpacs,amsmath,showkeys]{revtex4}

\usepackage[dvips]{graphicx}
\usepackage{epstopdf}
\usepackage{amssymb}
\usepackage{epstopdf}

\newcommand{\ba}{\begin{eqnarray}}
\newcommand{\ea}{\end{eqnarray}}
\newcommand{\bmath}{\begin{mathletters}}
\newcommand{\emath}{\end{mathletters}}
\newcommand{\ban}{\begin{eqnarray*}}
\newcommand{\ean}{\end{eqnarray*}}

\draft

\begin{document}

\title{U(3) and Pseudo-U(3) Symmetry of the Relativistic Harmonic Oscillator}
\author{Joseph N. Ginocchio}
\address{MS B283, Theoretical Division, Los Alamos National Laboratory, Los
Alamos, New Mexico 87545, USA}
\date{\today}

\begin{abstract}
We show that a Dirac Hamiltonian with equal scalar and vector harmonic oscillator potentials has not only a spin symmetry but an U(3) symmetry  and that a Dirac Hamiltonian with scalar and vector harmonic oscillator potentials equal  in magnitude but opposite in sign has not only a pseudospin symmetry but a pseudo-U(3) symmetry. We derive the generators of the symmetry for each case. 
\end {abstract}

\keywords{ Symmetry,
Dirac Hamiltonian, Relativistic mean field theory, Spin, Pseudospin, SU(3), pseudo-SU(3)}

\pacs{ 21.60.-n, 21.10.-k, 02.20.-a}

\maketitle

\newpage

As is well-known, the non-relativistic spherical harmonic oscillator has degeneracies in addtion to those due to rotational invariance. The energy spectrum depends only on the {\it total} harmonic oscillator quantum number $N$, $N= 2n + \ell$, where n is the radial quantum number and $ \ell$ is the orbital angular momentum. Hence the states with $\ell = N, N -2, \dots 0$ or $1$ have the same energy. These degeneracies are produced by an U(3) symmetry \cite {elliott} .  This U(3) symmetry has been influential in connecting the shell model with  collective motion \cite {igal}. Also the energy does not depend on the orientation of the spin and hence the non-relativistic harmonic oscillator has a spin symmetry as well. 

Since relativistic models of nuclei are now so prevalent \cite{ring}, we can ask if U(3) symmetry resides in the relativistic harmonic oscillator. 
Indeed the Dirac Hamiltonian, $H$,
for which the scalar, $V_S(\vec r)$,
and vector, $V_V(\vec r)$, potentials are equal and harmonic has been solved analytically and is invariant under a spin symmetry \cite {gino05,gino04}; that is $[\vec S,H]=0$ where $\vec S$ is given in Eq.(\ref {gen}). Just as for the non-relativistic harmonic oscillator, the spherically symmetric relativistic harmonic oscillator energy spectrum depends only on the total harmonic oscillator quantum number $N$, although the energy spectrum for the relativistic harmonic oscillator spectrum  in general does not have a linear dependence on $N$ as does the non-relativistic harmonic oscillator. This suggests that the relativistic  harmonic oscillator does have an U(3) symmetry. If this is the case, the question is: what are the relativistic generators? In this letter we shall show that there is indeed a U(3) symmetry and we shall derive the generators.

The Dirac Hamiltonian for a spherical harmonic oscillator with spin symmetry is

\begin{equation}
H ={\vec \alpha}\cdot {\vec p}
+ \beta M + (1+\beta)V( r),
\label {dirac}
\end{equation}
where ${\vec \alpha}$, $\beta$ are the Dirac matrices,  ${\vec p}$ is the momentum,  $M$ is the mass, $V(r) ={{M\omega^2}\over 2}\ r^2$,  ${\vec r}$ is the radial coordinate, $r$ its magnitude, and the velocity of light is set equal to unity, c =1.
The generators for the spin SU(2) algebra and the orbital angular momentum SU(2) algebra,
${{\vec S}},{\vec L}$,
which commute with the
Dirac Hamiltonian, $[\,H\,,\, {\vec S}\,] = [\,H\,,\,{\vec L}\,] =0$, 
are given by \cite {bell}
\begin{equation}
{\vec S} =
\left (
\begin{array}{cc}
 {\vec s}   & 0 \\
0 & U_p\ {\vec  s}\ U_p
\end{array}
\right ), \ 
{\vec L} = \left (
\begin{array}{cc}
{\vec { \ell}}   & 0 \\
0 & U_p\  {\vec { \ell}}\  U_p
\end{array}
\right ),
\label{gen}
\end{equation}
where
$ {\vec s} = {\vec \sigma}/2$ are the usual spin generators,
${\vec \sigma}$ the Pauli matrices, $ {\vec { \ell}} = {({\vec r }\times {\vec p})\over \hbar}$, and
$U_p = \, {\mbox{\boldmath ${\vec \sigma}\cdot {\vec p}$} \over p}$ is the
helicity unitary operator introduced in \cite {draayer}.

The non-relativistic U(3) generators are  the orbital angular momentum $ {\vec { \ell}}$, the quadrupole operator $q_m = {1\over  \hbar M\omega}{\sqrt{3\over 2} } (M^2 {\omega}^2[rr]_m^{(2)}  + {[pp]_m^{(2)}}$),
where $[rr]_m^{(2)}$ means coupled to angular momentum rank 2 and projection $m$, and ${\cal N}_{NR} = {1\over 2\sqrt{ 2}  \hbar M\omega  } (M^2 {\omega}^2r^2  + p^2)- {3\over 2}$. They form the closed U(3) algebra 
\begin{subequations}
\begin{equation}
[{ \cal N}_{NR},{\vec { \ell}}] = [ {\cal N}_{NR},q] = 0, 
\end{equation}
\begin{equation}
[ {\vec { \ell}},{\vec { \ell}}]^{(t)} = - \sqrt{2}\  {\vec { \ell}}\ {\delta}_{t,1} ,\  [ {\vec { \ell}},q]^{(t)} = -\sqrt{6}\  {q}\ {\delta}_{t,2}, \ [q,q]^{(t)} =  3\sqrt{10}\ {\vec { \ell}}\ \delta_{t,1},
\end{equation}
\label {cr}
\end{subequations}
with $ {\cal N}_{NR}$ generating a U(1) algebra whose eigenvalues are the total number of quanta $N$ and ${\vec { \ell}}, q$ generating an SU(3) algebra.
In the above we use the coupled commutation relation between two tenors, $T_1^ {(t_1)},T_2^ {(t_2)}$ of rank $t_1,t_2$ which is  $[T_1^ {(t_1)},T_2^ {(t_2)}]^{(t)}=  [T_1^ {(t_1)}T_2^ {(t_2)}]^{(t)}- (-1)^{t_1+t_2-t}[T_2^ {(t_2)}T_1^ {(t_1)}]^{(t)}$ \cite{french}.

The relativistic orbital angular momentum generators ${\vec L}$ are given in Eq. (\ref{gen}).  We shall now determine the the quadrupole operator $ Q_m$ and monopole operator $  {\cal N}$ that commute with the Hamiltonian in Eq.(\ref {dirac}).  
In order for the quadrupole generator 
\begin{equation}
{ Q_m} =
\left (
\begin{array}{cc}
 {(Q_m)_{11}}   &  {(Q_m)_{12}} {\vec \sigma}\cdot {\vec p}\\
 {\vec \sigma}\cdot {\vec p}\ {(Q_m)_{21}}  &  {\vec \sigma}\cdot {\vec p}\  {(Q_m)_{22}}\   {\vec \sigma}\cdot {\vec p}
\end{array}
\right ), 
\label{qgen}
\end{equation}
to commute with the Hamiltonian, $[Q_m,H] = 0$, the matrix elements must satisfy the 
conditions,

\begin{subequations}
 \begin{equation}
 {(Q_m)_{12}} = {(Q_m)_{21}},
 \end{equation}
 \begin{equation}
[{( Q_m)_{11}} ,V]+[{(Q_m)_{12}},p^2] = 0,
 \end{equation}
\begin{equation}
[{( Q_m)_{12}} ,V]+[{(Q_m)_{22}},p^2] = 0,
 \end{equation}
 \begin{equation}
 {(Q_m)_{11}}= {(Q_m)_{12}}\ (V+2M) + {(Q_m)_{22}}\  p^2.
 \end{equation}
 \label{qme}
\end{subequations}

One solution is
\begin{eqnarray}
{ Q_m} =
\lambda_2\ \left (
\begin{array}{cc}
 {{M\omega^2}\over 2}\  ({{M\omega^2}\over 2}\ r^2+2M) [rr]^{(2)}_m\ + [pp]^{(2)}_m&\ \ \  {{M\omega^2}\over 2}\ [rr]^{(2)}_m\   {\vec \sigma}\cdot {\vec p}\\\
{\vec \sigma}\cdot {\vec p}\ {{M\omega^2}\over 2}\ [rr]^{(2)}_m  & \ \ \ [pp]^{(2)}_m
\end{array}
\right ), 
\label{Q}
\end{eqnarray}
where $\lambda_2$ is an overall constant undetermined by the commutation of $Q_m$ with the Dirac Hamiltonian. 

For this quadrupole operator to form a closed algebra, the commutation with itself must be the orbital angular momentum operator as in Eq. (\ref{cr}). This commutation relation gives
\begin{equation}
[Q,Q]^{(t)} =\sqrt{10}\  \lambda_2^2\ {{M\omega^2}}{\hbar}^2\ \left (
\begin{array}{cc}
 ({{M\omega^2}\over 2}\ r^2+2M) \ {\vec \ell} & \  \ {\vec \ell}\  {\vec \sigma}\cdot {\vec p}\\
{\vec \sigma}\cdot {\vec p}\ {\vec \ell} & \  \ 0
\end{array}\right )= \sqrt{10}\  \lambda_2^2\ {{M\omega^2}}{\hbar}^2\\ (H +M)\ {\vec L}\  \delta_{t,1},
 \label {crq}
\end{equation}
and we get the desired result if $\lambda_2 =  \sqrt{ 3 \over {{M\omega^2}} {\hbar}^2(H+M)}$.
The quadrupole operator then becomes

\begin{eqnarray}
{ Q_m} =
  \sqrt{ 3 \over {{M\omega^2}} {\hbar}^2(H+M)}\ \left (
\begin{array}{cc}
 {{M\omega^2}\over 2}\  ({{M\omega^2}\over 2}\ r^2+2M) [rr]^{(2)}_m\ + [pp]^{(2)}_m& \ \ \ {{M\omega^2}\over 2}\ [rr]^{(2)}_m\ {\vec \sigma}\cdot {\vec p}\\\
{\vec \sigma}\cdot {\vec p}\ {{M\omega^2}\over 2}\ [rr]^{(2)}_m  &\ \ \  [pp]^{(2)}_m
\end{array}
\right ).
\label{Q}
\end{eqnarray}

In order for the monopole  generator 
\begin{equation}
{  {\cal N}} =
\left (
\begin{array}{cc}
 {( {\cal N})_{11}}   &  {( {\cal N})_{12}} {\vec \sigma}\cdot {\vec p}\\
 {\vec \sigma}\cdot {\vec p}\ {( {\cal N})_{21}}  &  {\vec \sigma}\cdot {\vec p}\  {( {\cal N})_{22}}\   {\vec \sigma}\cdot {\vec p}
\end{array} 
\right )+ {\cal N}_0, 
\label{qgen}
\end{equation}
to commute with the Hamiltonian, $[ {\cal N},H] = 0$, the matrix elements must satisfy the 
conditions in Eq.(\ref {qme}) with $Q_m$ replaced by $ {\cal N}$. $ {\cal N}_0$ is a constant. A solution is
\begin{eqnarray}
{ {\cal N}} =
\lambda_0\ \left (
\begin{array}{cc}
 {{M\omega^2}\over 2}\  ({{M\omega^2}\over 2}\ r^2+2M) r^2\ + p^2&\ \ \  {{M\omega^2}\over 2}\ r^2\   {\vec \sigma}\cdot {\vec p}\\\
{\vec \sigma}\cdot {\vec p}\ {{M\omega^2}\over 2}\  r^2  & \ \ \ p^2
\end{array}
\right )+ {\cal N}_0.
\label{N}
\end{eqnarray}

Straightfoward calculations show that  ${\cal N}$ commutes with the the other generators as well as the Dirac Hamiltonian and consequently  is the U(1) generator. However, the  constants $\lambda_0,  {\cal N}_0$ are undetermined by these commutation relations. These constants are determined instead by requiring that the eigenvalue of ${\cal N}$ is the the total harmonic oscillator number, $N$; that is, ${\cal N}\ \Psi_N= N\ \Psi_N$, where $\Psi_N$ are the eigenfunctions of the Dirac Hamiltonian, $H\ \Psi_N= E_N\ \Psi_N$. Using the facts that \cite {gino04, gino05}

\begin{equation}
 {\Psi}_N =
\left (
\begin{array}{c}
 g  \\
 {  {\vec \sigma}\cdot {\vec p}\over {E_N +M}}\  g
 \end{array} 
\right ), [p^2 +(E_N+M)V(r) -2\ \hbar \sqrt{(E_N+M)M\omega^2}(N+{3\over2})]g = 0,
\label{L0}
\end{equation}
we derive that 
\begin{equation}
{\cal N} {\Psi}_N =
 [2\ \hbar\ \lambda_0\ \sqrt{(E_N+M)M\omega^2}(N+{3\over2})+{\cal N}_0]{\Psi}_N = N\ {\Psi}_N,
\label{NC}
\end{equation}
which determines $\lambda_0={1\over 2\  \hbar \sqrt{(H+M)M\omega^2}}, {\cal N}_0 = -{3\over2}$.

In the non-relativisitc limit, $H \rightarrow M, M \rightarrow \infty$,

\begin{eqnarray}
{ Q_m} \rightarrow
 {1\over  \hbar M\omega}{\sqrt{3\over 2} }\ \left (
\begin{array}{cc}
 {{M^2{\omega}^2}}\   [rr]^{(2)}_m\ +{ [pp]^{(2)}_m }& 0\\
0 & 0
\end{array}
\right )
=
  \left (
\begin{array}{cc}
q_m & 0\\
0  & 0
\end{array}
\right ),\\ { \cal N} \rightarrow
 {1\over 2\ \sqrt{ 2} \hbar M\omega} \ \left (
\begin{array}{cc}
 {M^2{\omega}^2\   r^2\ +p^2 }& 0\\
0 & 0
\end{array}
\right )- {3\over 2}
=
  \left (
\begin{array}{cc}
{ \cal N}_{NR} & 0\\
0  & - {3\over 2}
\end{array}
\right ),
\label{Qq}
\end{eqnarray}
which agrees with the non-relativistic generators.

The commutation relations are then those of the U(3) algebra, 
\begin{subequations}
\begin{equation}
[{ \cal N},{\vec { L}}] = [ {\cal N},Q_m] = 0. 
\end{equation}
\begin{equation}
[ {\vec { L}},{\vec { L}}]^{(t)} = -\sqrt{2}\  {\vec { L}}\ {\delta}_{t,1} ,\  [ {\vec { L}},Q]^{(t)} =   -\sqrt{6}\  {Q}\ {\delta}_{t,2}, \ [Q,Q]^{(t)} = 3\sqrt{10}\ {\vec { L}}\ \delta_{t,1}.
 \label {crr}
\end{equation}
\end{subequations}
The spin generators in Eq.(\ref{gen}), $\vec S$, commute with the U(3) generators as well as the Dirac Hamiltonian, and so the invariance group is U(3)$ \times $SU(2), where the SU(2) is generated by the spin generators, $[{\vec S},{\vec S}]^{(t)} = - \sqrt{2}\ {\vec S}\ \delta_{t,1}$.

The Dirac Hamiltonian for a spherical harmonic oscillator with pseudospin symmetry is \cite {gino97}

\begin{equation}
{\tilde H }={\vec \alpha}\cdot {\vec p}
+ \beta M + (1-\beta)V( r),
\label {pdirac}
\end{equation}
which explains the pseudospin doublets observed in nuclei \cite {gino05}. This pseudospin Hamiltonian can be obtained from the spin Hamiltonian with a transformation by $\gamma_5 = \left (
\begin{array}{cc}
0  & 1\\
1 & 0
\end{array}
\right ) $ and $M \rightarrow -M$, but $M\omega^2 \rightarrow M\omega^2$.
The generators for the pseudospin \cite {ami} and pseudo- U(3) algebra which commute with the
Dirac Hamiltonian, $[{\tilde H}, {\vec {\tilde S} }] = [{\tilde H}, {\vec {\tilde L} }] =[{\tilde H},{\tilde{Q}_m}] = [{\tilde H},{\tilde{\cal N}}]=0$, 
are then obtained by the same transformation and are given by 
\begin{subequations}
 \begin{equation}
 {\vec {\tilde S} }=
\left (
\begin{array}{cc}
 U_p\ {\vec  s}\ U_p   & 0 \\
0 & {\vec s}
\end{array}
\right ), \ 
{\vec {\tilde L}} = \left (
\begin{array}{cc}
U_p\  {\vec { \ell}}\  U_p  & 0 \\
0 & {\vec { \ell}} 
\end{array}
\right ),
\end{equation}
\begin{equation}
{\tilde Q}_m= \sqrt{ 3 \over {{M\omega^2}} {\hbar}^2({\tilde H}-M)}\ \left (
\begin{array}{cc}
 [pp]^{(2)}_m& {\vec \sigma}\cdot {\vec p}\ {{M\omega^2}\over 2}\ [rr]^{(2)}_m \ \ \ \\\
{{M\omega^2}\over 2}\ [rr]^{(2)}_m\ {\vec \sigma}\cdot {\vec p} &\ \ \  {{M\omega^2}\over 2}\  ({{M\omega^2}\over 2}\ r^2-2M) [rr]^{(2)}_m\ + [pp]^{(2)}_m
\end{array}
\right ),
\end{equation}
\begin{equation}
{\tilde {\cal N}} =
{1\over 2\  \hbar \sqrt{M\omega^2({\tilde H}-M)}}\ \left (
\begin{array}{cc}
 p^2&\ \ \ {\vec \sigma}\cdot {\vec p} \ {{M\omega^2}\over {2}}\ r^2 \\
{{M\omega^2}\over 2}\   {\vec \sigma}\cdot {\vec p}\  r^2  & \ \ \ {{M\omega^2}\over 2}\  ({{M\omega^2}\over 2}\ r^2-2M) r^2\ + p^2
\end{array}
\right )- {3\over 2}.
\end{equation}
\label{pgen}
\end{subequations}

The relativistic non-spherical  harmonic oscillator has also been solved analytically \cite{gino04} as well. The relativistic axially symmetric deformed  harmonic oscillator will have a U(2) symmetry as  does the non-relativistic axially symmetric deformed harmonic oscillator. This will be discussed in a forthcoming paper.  

In summary, we have shown that a Dirac Hamiltonian with equal scalar and vector harmonic oscillator potentials has an U(3)$\times$ SU(2) symmetry  and that a Dirac Hamiltonian with scalar and vector harmonic oscillator potentials equal  in magnitude but opposite in sign has a pseudo-U(3) $\times$ pseudo-SU(2) symmetry and we have derived the corresponding generators for each case. If speculation that an anti-nucleon can be bound inside a nucleus is valid \cite {thomas}, the anti-nucleon spectrum will have an approximate spin symmetry and, most likely an approximate U(3) symmetry, because the vector and scalar potentials are approximately equal and are very strong \cite {gino05}.

\end{document}